\documentstyle[12pt,aaspp4]{article}
\slugcomment{}

\lefthead{Sankrit, Hester}
\righthead{Modeling Blister HII Regions}

\begin{document}

\title{Modeling the Photoionized Interface in Blister HII Regions}

\author{Ravi Sankrit\footnotemark[1]}
\affil{The Johns Hopkins University}
\authoraddr{Rm 366 Bloomberg, 3400 N. Charles St., Baltimore, MD 21218}
\and
\author{J. Jeff Hester\footnotemark[2]}
\affil{Arizona State University}
\authoraddr{Box 871504, Arizona State University, Tempe, AZ 85287}

\footnotetext[1]{ravi@pha.jhu.edu}
\footnotetext[2]{jhester@asu.edu}

\begin{abstract}

We present a grid of photoionization models for the emission from
photoevaporative interfaces between the ionized gas and molecular cloud
in blister H~II regions.  For the density profiles of the emitting gas
in the models, we use a general power law form calculated for
photoionized, photoevaporative flows by Bertoldi (1989).  We find that
the spatial emission line profiles are dependent on the incident flux,
the shape of the ionizing continuum and the elemental abundances.  In
particular, we find that the peak emissivity of the [S~II] and [N~II]
lines are more sensitive to the elemental abundances than are the total
line intensities.  The diagnostics obtained from the grid of models can
be used in conjunction with high spatial resolution data to infer the
properties of ionized interfaces in blister H~II regions.  As an
example, we consider a location at the tip of an ``elephant trunk''
structure in M~16 (the Eagle Nebula) and show how narrow band HST-WFPC2
images constrain the H~II region properties.  We present a
photoionization model that explains the ionization structure and
emission from the interface seen in these high spatial resolution
data.

\end{abstract}


\section{Introduction}

A blister H~II region is formed when a massive star is born near the
edge of a molecular cloud.  The radiation from the newborn star rapidly
ionizes the surrounding material and the ionization front breaks
through the surface of the cloud creating a cavity.  The photoionized
gas in the cavity is exposed to view and can be observed as a blister
on the surface of the molecular cloud.  The concept of a blister was
introduced by \cite{zuc73} to describe the Orion Nebula, one of the
nearest and definitely the best studied H~II region.  The term
``blister model'' was introduced by \cite{isr78} who studied a sample
of about 30 galactic H~II regions and, based on their positional
relationship to CO emitting molecular clouds, concluded that almost
every optically observable H~II region is a blister H~II region.

Blister H~II regions serve as useful probes of current interstellar
abundances in the Galaxy, since the stars responsible for maintaining
the ionization are less than about 10 million years old.  Classical
techniques of nebular analysis (described in several textbooks, such as
\cite{all84} and \cite{ost89}), based mainly on optical spectra, have
been used to determine abundances in several H~II regions and map
abundances in the Galaxy (e.g.~\cite{haw78}, \cite{sha83}).  These
methods have two well known uncertainties.  First, lines from all the
ionization stages of an element present in an H~II region are usually
not observed.  This requires prescriptions to correct for unseen stages
of one element based on ionic ratios of a different element.  Second,
temperatures derived from lines of one species (such as [O~III]) are
assumed when obtaining the abundance of another species (such as
[S~II]) which is formed in a different zone.  These uncertainties are
discussed by \cite{fre81}.

An alternative method of determining abundances is by using
photoionization models to interpret H~II region spectra.  This approach
has been applied in some detail to the Orion Nebula (\cite{bal91},
\cite{rub91}) and has the advantage that the electron temperature and
ionization structure are calculated self-consistently.  Furthermore, it
becomes unnecessary to implement ad-hoc ionization correction
schemes.  However a major limitation of this method (as emphasized by
\cite{rub98}) is that the input density and geometry need to be
specified in order to calculate models, and these in general are not
well known.

In a blister H~II region, the density profile of the interface between
the molecular cloud and the ionized gas is determined by the ionizing
radiation driving a photoevaporative flow off the surface of the
molecular cloud.  The sharply stratified ionization structure of this
interface was observed in the Orion Nebula by \cite{hesetal91} in
\textit{Hubble Space Telescope} (HST) \textit{Wide Field Camera}
narrow band images.  The ionization stratification was discussed in the
broader context of interpreting H~II region spectra by \cite{hes91},
who also suggested that most of the [S~II] emission arises in a very
narrow transition zone between the H~II region interior and the
photodissociation region of the molecular cloud.  The transition zone
occurs beyond the hydrogen ionization edge, where photons with energies
higher than 10.4 eV (the ionization potential of S$^{0}$) keep the
sulphur singly ionized.  More recently, \cite{hes96} presented HST
\textit{Wide Field and Planetary Camera - 2} (WFPC2) images of the
``elephant trunk'' structures in the blister H~II region M~16.  In
these images, the photoionized interface is seen in tangency and the
ionization stratification is clearly resolved.  The [S~II] emitting
zone is very narrow -- it has a width of about $8 \times 10^{15}$~cm
($\lesssim$~0\farcs3 at the assumed distance of 2000 pc).  In that
work, we used an empirical density profile derived from the H$\alpha$
emission profile and presented a photoionization model that
successfully reproduced the main features of the emission from the
interface.

In this paper, we use photoionization models in conjunction with high
spatial resolution data to develop a framework for
interpreting H~II region spectra.  Arbitrary assumptions about the
density structure are not made, since the models are constrained by the
structure observed in the high resolution images.  Once such a
framework is established, it will be a powerful method for obtaining
the physical properties of the emitting gas in H~II regions.

We first present a grid of models for the interface in blister H~II
regions using the density profiles calculated by \cite{ber89} for
photoionized, photoevaporative flows.  (These calculations are a
significant improvement over earlier work because they allow for an
ionization front of finite width, rather than treating it as a
discontinuity, and they treat non-equilibrium ionization and energy
deposition).  We vary the incident stellar continuum, the ionizing flux
and the elemental abundances and examine the dependence of nebular
properties on these input parameters.  We focus on the diagnostics
provided by spatially resolved line strengths (such as can be obtained
by HST for several nearby H~II regions).  We then consider the M~16
data in detail and present a photoionization model for the emission.
We compare the model results to published spectra and show how
knowledge of the structure of the emitting region is crucial for
interpreting ground based observations.  We conclude with a summary and
a consideration of the implications of our work, and future
directions.

\section{Photoionization Models}

\subsection{Input Parameters}

The basic input parameters for the photoionization models are the shape
and intensity of the incident ionizing continuum and the elemental
abundances.  The interface between the molecular cloud and the H~II
region interior is the result of a photoevaporative flow driven by the
incident stellar continuum (which is also responsible for ionizing the
gas).  The solutions for such a flow determine the density profile of
the interface, which also needs to be specified.

Hydrodynamic models for photoionized, photoevaporative flows off
spherical cloud surfaces have been calculated by \cite{ber89} and
\cite{ber96}.  The form of the density profile they find may be written
as follows:
\begin{equation}
\label{EQD} n(x) =  n_{0}(1 + \case{R-x}{r_{c}})^{-2.5}
\end{equation}
Here, $x$ is the distance from the ionizing source, $R$ is the distance
from the source to the ionization edge, and $n_{0}$ is the number
density at $x = R$.  (Note that the density is a function of the
distance from the interface, $R~-~x$, and the actual distance from the
source is not explicitly required in the models).  The radius of
curvature of the evaporating surface is given by $r_{c}$ and is
effectively the scale length for the flow.  The values of $R$ and
$n_{0}$ depend on the ionizing continuum.  For each of our models (see
below), we start by performing a calculation with
$R~=~5\times10^{17}$~cm and $n_{0}~=~1500$~cm$^{-3}$, and then iterate
using the predicted values at the ionization edge till we obtain the
correct values.  (We found that 3 iterations were sufficient in every
case).  We keep the value of $r_{c}$ fixed at $2\times10^{17}$~cm for
all the models, anticipating our analysis of the M~16 data in \S3.

The validity of using equation \ref{EQD} for the density profile is subject to
the following caveat.  The models calculated by \cite{ber89} did not
include the effects of dust on the flow.  It was shown by \cite{bal91}
that grains absorb, and are ionized by the incident continuum and cause
a radial acceleration of the gas away from the star.  This phenomenon
will affect the outer regions of the flow and alter the density
profile.  In order to asses the effect on the density profile at and
near the interface (which we are concerned with here), the hydrodynamic
and photoionization problem would have to be solved self consistently,
taking into account the presence of dust - a task beyond the scope of
this paper.

We calculate a grid of 60 models, using all combinations drawn from 4
stellar atmospheres, 3 values for the ionizing flux and 5 sets of
elemental abundances.  The stellar atmospheres used are PHOENIX models
of B0 (T$_{eff}$ = 33,340~K), O8 (T$_{eff}$ = 38,450~K), O6 (T$_{eff}$
= 43,560~K), and O4 (T$_{eff}$ = 48,670~K) stars, all with gravities,
log(g) = 3.9 (\cite{hau97}, Aufdenberg, private communication).
PHOENIX models represent an advance over earlier model stellar
atmospheres since they use spherical geometry, calculate the strongest
lines of important elements in non-LTE, and include line blanketing.
In Figure \ref{CONTINUA} these model spectra are plotted, normalized to
their respective peak values.  Also shown on the plot are the
ionization potentials of S$^{0}$ (10.36 eV) and O$^{+}$ (35.1 eV).  The
values used for the incident ionizing flux are log [$\Phi$(H)] = 10.5,
11.0 and 11.5, where where $\Phi$(H) is the number of ionizing photons
(energy $\geq$ 13.6 eV) per second per cm$^2$ incident on the gas.  In
all the models we keep the helium abundance at solar.  The heavier
element abundances are varied simultaneously and values of 0.2, 0.4,
0.6, 0.8 and 1.0 times solar (where solar values are taken from
\cite{gre89}) are used in the grid of models.  We use the publicly
available code, CLOUDY (\cite{fer98}) to calculate the photoionization
models.

One concern about the modeling procedure is that we are using an
equilibrium code (CLOUDY) to model the emission from a photoevaporative
flow where non-equilibrium effects may be important.  
At the interface, the cooling times are short (a few tens of years)
compared to the dynamic time scales (a few thousand years), so
thermal equilibrium is a valid assumption.  However sufficiently
close to the ionization front, photoionization equilibrium will not
hold.  The effect of non-equilibrium on the predicted emission
lines was studied by \cite{har77}.  He found that the ionization
fraction differed only moderately between equilibrium and non-equilibrium
models, and that happened within about 10$^{15}$~cm of the front.  He also
found that the lines strengths most affected were [N~I] $\lambda$5199
and [O~I] $\lambda$6300 which varied about 40\% and 15\% respectively.
Both these lines are formed beyond and are narrower than the [S~II] zone.
While it would be useful to examine the effect of photoionization 
non-equilibrium in detail for a wider range of conditions than was done
by \cite{har77}, we expect our models to be suitable representation 
of reality.

\subsection{Results}

The photoionization models predict the highly stratified ionization
structure of the interface.  The H$\alpha$ more or less follows the
density profile as expected.  The [S~II] comes from a narrow zone near
the interface that peaks sharply just beyond the H$\alpha$ peak, in a
region where the ionization fraction of hydrogen is rapidly falling
and sulphur (whose ionization potential is lower than hydrogen) is
kept singly ionized.  The [O~III] emission profile is always less
peaked and often quite extended compared to the H$\alpha$ and [S~II].
While all models show these general characteristics of the low
ionization and high ionization gas, there are significant differences
in the details of the spatial profiles and line strengths among the
various models.  We now turn to a discussion of these differences and
their dependence on input parameters.

In Figure \ref{CPHI} we show the density and the H$\alpha$,
[S~II]~$\lambda\lambda$6716,6731 and [O~III]~$\lambda$5007 emissivity
as a function of distance from the interface for two models.  Both
models use an O6 star continuum and abundances of 0.6 solar for
elements heavier than helium.  The models shown differ in the incident
ionizing fluxes -- $\Phi_{\rm{H}} =
10^{10.5}$~photons~s$^{-1}$~cm$^{-2}$ (top panel) and $\Phi_{\rm{H}} =
10^{11.5}$~photons~s$^{-1}$~cm$^{-2}$ (bottom panel).  The most
dramatic difference between the two models is the density at the
ionization front, which is about 3 times higher for the higher incident
flux.  Consequently the peak H$\alpha$ emission is higher by about a
factor of 10 (since emissivity is proportional to the square of the
density).  This is a result of the ionization front being pushed into a
cloud where the density increases towards the interior -- a larger flux
of ionizing photons will push the front further up the density ramp.
The [S~II] and [O~III] emissivities are also correspondingly higher.

In both plots of Figure \ref{CPHI}, the origin of the x-axis is chosen
to be at the peak of the [S~II] emission.  We can see from these plots,
that the width of the [S~II] zone varies considerably with changes in
the incident ionizing flux.  If we define the width as being the
distance at which the [S~II] emissivity falls to half its peak value,
then the widths for the two cases shown are $11.2\times10^{15}$~cm for
the lower ionizing flux (top panel) and $4.4\times10^{15}$~cm for the
higher ionizing flux (bottom panel).  At a distance of 1~kpc, these
widths correspond to 0\farcs75 and 0\farcs29 respectively.

In Figure \ref{CSTAR}, models with different shapes for the incident
continuum are shown.  The top and bottom panels show models using O8
and O4 stellar atmospheres.  Both models use an incident ionizing flux
of $10^{11.0}$~photons~s$^{-1}$~cm$^{-2}$ and metal abundances of 0.6
times solar.  The [O~III] emission is higher by a factor of about 6 for
the O4 model while H$\alpha$ and [S~II] are about the same.  This is
because the fraction of photons produced which are capable of ionizing
oxygen to O$^{++}$ is much lower for O8 stars than for O4 stars (Figure
\ref{CONTINUA}).  Here again, the [S~II] profiles are significantly
different.  The O8 star produces an [S~II] zone that is
$\sim~12.3\times10^{15}$~cm wide (top panel) and the hotter O4 star
produces a narrower [S~II] zone ($\sim~6.4\times10^{15}$~cm).

The variation of emission profiles with abundance is shown in Figure
\ref{CABUND}.  Both models use an O6 stellar atmosphere and an ionizing
flux of $10^{11.0}$~photons~s$^{-1}$~cm$^{-2}$.  The models shown in
the top and bottom panels have the abundances of all elements heavier
than helium set to 0.2 times solar and 0.8 times solar respectively.
The H$\alpha$ intensity is about the same in both cases, and the total
[S~II] intensity is 1.6 times higher in the model with the higher
abundance.  The ratio of the peak [S~II] emissivity to the peak
H$\alpha$ emissivity shows a stronger contrast between the models.  In
the case of the low abundance model this ratio is 0.27,  while for the
high abundance model it is 0.55, which is a factor of 2 higher.  

\subsubsection{Stellar Atmosphere Models}

The spectral energy distribution of the incident ionizing continuum is
a key ingredient in HII region models.  Since the ionizing photons
cannot be observed for any but a handful of stars, we depend on
continua predicted by theoretically computed stellar atmospheres.  In
this paper we are using atmospheres calculated by the PHOENIX code,
which treats the atmosphere as spherically symmetric, includes line
blanketing, and treats the strongest lines of several important
elements in non-LTE.  These models are an improvement over the widely
used ATLAS models (\cite{kur91}) which are plane-parallel and neglect
non-LTE effects entirely.

Another set of theoretical non-LTE stellar atmospheres, the CoStar
Models (\cite{sch97}) have also been used recently in H~II region
models (\cite{sta97}).  CoStar and PHOENIX models differ in significant
ways.  CoStar models include the effects of stellar winds on the 
emitted continuum, while the PHOENIX models are spherically extended but
hydrostatic atmospheres.  However, CoStar models do not include line
blanketing in the calculation of the temperature structure whereas PHOENIX
models include line blanketing in a self consistent way in determining
the temperature structure of the atmosphere.  Furthermore CoStar models
treat only H and He in non-LTE while PHOENIX treats over 50,000 of the
strongest lines of several elements in non-LTE.  In this section, we
examine the effect of using different model atmospheres on the
predicted H~II region photoionization model.

Figure \ref{COS_PHX_CONT} shows the spectral energy distributions of
the PHOENIX O4V model (T$_{*}$ = 48670 K) and the CoStar E2 model
(T$_{*}$ = 48500 K).  Also shown are the S$^{0}$ and O$^{++}$
ionization edges.  The CoStar spectrum is flatter and has a greater
flux at higher energies than the PHOENIX spectrum.  This effect is due
to the wind and is seen in \cite{sta97} where CoStar and ATLAS spectra
are compared.  In order to understand the effects of using different
stellar atmospheres on H~II region spectra, we calculated a
photoionization model using the CoStar spectrum along with an incident
ionizing flux of $10^{11.0}$~photons~s$^{-1}$~cm$^{-2}$, and metal
abundances of 0.6 times solar.  In Figure \ref{COS_PHX_SPEC} we compare
this CoStar model with a model using the PHOENIX O4V ionizing continuum
with the same incident flux and abundances.  The difference between
these models is significant - the CoStar continuum produces twice as
much [O~III] as the PHOENIX continuum.  The CoStar model also produces
more [S~II] and has a somewhat broader [S~II] zone.  The peak H$\alpha$
is lower in the CoStar model, so the peak [S~II] to peak H$\alpha$ is
higher by a factor of 2 compared to the PHOENIX model.

We wish to caution the reader that currently it is very uncertain as to
which of these models (PHOENIX or CoStar) is closer to reality.
However our comparison has shown that in their current forms, the
ionizing continua from by PHOENIX and CoStar predict very different
H~II region line strengths.  A thorough understanding of H~II region
spectra can be of value in validating atmosphere models as pointed out
by \cite{rub95}.  It is beyond the scope of this work to pursue this
issue further and for the rest of the paper, we will discuss models
using only PHOENIX stellar continua.

\subsection{Diagnostics}

The grid of models we have calculated predict several quantities that
could serve as useful starting points in analyzing spectra of blister
H~II regions, particularly for the cases where the emitting interface
is spatially resolved.  We now present and discuss some of these
diagnostics.

In most of the following plots, we will be visualizing the data in a
somewhat specialized way which is worth describing here.
The quantity of interest is plotted on the y-axis.  The x-axis
is simply the model number (from 1 to 60, with the tick marks
suppressed).  The 4 large divisions in the plot correspond to stellar
atmospheres used in the model (models 1 through 15 use a B0 atmosphere
and so on).  Within each set of 15 models, groups of 5 (differentiated
by the symbols used) correspond to different values for the ionizing
flux.  Finally, within each group of 5, the heavy element abundances go
from 0.2 solar to 1.0 solar in steps of 0.2 with increasing model
number.

In the top panel of Figure \ref{HADEN} we show the maximum density
reached in each model.  This is the hydrogen density at the ionization
edge defined (for convenience) to be the point where $n_{e}$/$n_{H}$
has fallen below 0.001. (We note, however that the observed lines are
not emitted so far into the photodissociation region -- for instance,
the [S~II] peak occurs where $n_{e} \sim 0.7n_{H}$).  The density at
the ionization edge depends strongly on the ionizing flux (for the
reason mentioned in the previous section).  There is, however no
dependence on the stellar temperature; (note that the four large
divisions in the plot correspond to different ionizing stars).  For
higher incident flux, there is a noticeable dependence on the
abundance.  A higher abundance of metals implies that more atoms
compete for the ionizing photons and the maximum density reached is
lower.  The peak H$\alpha$ emission depends on the density at the
ionization edge.  The bottom panel of Figure \ref{HADEN} is a scatter
plot of the peak H$\alpha$ against the maximum density reached in the
models.  As expected there is a strong correlation between the
H$\alpha$ peak emissivity and the maximum density.  This in turn
implies that the peak H$\alpha$ emissivity is a good discriminant for
the incident ionizing flux.

Next, we discuss the peak emissivities and total line intensities of
three important diagnostic lines in H~II regions -- 
[O~III]~$\lambda$5007, [S~II]~$\lambda\lambda$6716,6731 and 
[N~II]~$\lambda$6584.  In what follows, all peak emissivities and total
intensities are taken relative to H$\alpha$, and this is to be
understood even if not explicitly mentioned.  Note that the peak
emissivity for different lines occurs at different locations.

Figure \ref{OIII} shows plots of the peak [O~III] emissivity (top
panel) and the total [O~III] intensity (bottom panel) for the grid of
models.  Both these quantities are most sensitive to the stellar type.
[O~III]~$\lambda$5007 is a high ionization line and in spectra of H~II
regions, it is an indicator of the hardness of the ionizing spectrum.
For a given ionizing flux, either the peak [O~III] emissivity or the
total [O~III] intensity could be used to distinguish the shape of the
incident ionizing continuum.  For models using a given star type and
ionizing flux, the [O~III] peak emissivity and intensities reach a
maximum for abundances about 0.4 solar.  The decrease in emission for
lower abundances is simply because there is less oxygen.  For higher
abundances, the increased oxygen abundance leads to the [O~III] 52
$\mu$m and 88 $\mu$m infrared lines dominating the cooling and lowering
the electron temperature of the gas (see e.g.~\cite{hen93}).  At lower
temperatures the intensity of [O~III], a collisionally excited line,
decreases.

In contrast to [O~III], the [S~II]~6716,6731~\AA\ lines trace low
ionization gas.  These [S~II] lines are the only strong optical lines
for any ionization stage of sulphur and are therefore have been of
great importance in estimating S abundances in nebulae.  However the
dominant state of sulphur in most H~II regions is S$^{++}$ and
obtaining only the S$^{+}$ abundance leads to large uncertainties in
the total sulphur abundance estimates (e.g.~\cite{den83}).  In Figure
\ref{SII} we show plots of the peak [S~II] emissivity (top panel) and
the total [S~II] intensity (bottom panel) for the grid of models.
(Note that the y-axis scales are different on the two plots).  For
models with the same ionizing continuum and flux, the peak [S~II]
emissivity is more sensitive than the total [S~II] intensity to changes
in abundance,  at least for lower abundances.  The sensitivity gets
better when the ionizing continuum is from hotter stars (O4 and O6 in
our grid).  Conversely, the peak [S~II] is less sensitive than the
total [S~II] intensity to the incident ionizing flux.  (Both quantities
decrease with increased ionizing flux).  Therefore, with spatially
resolved data for an H~II region interface, the peak [S~II] emissivity
can be used to estimate the sulphur abundance.

The ionization potential of N$^{0}$ is 14.5 eV, somewhat higher than
that of hydrogen.  In addition to direct ionization from the ground
state, N$^{0}$ in the excited \ $^{2}$D state can be ionized by photons
below the Lyman limit, and in some cases the latter may dominate the
ionization balance, and affect the [N~II] emission lines.  However in
all our models, the [N~II] $\lambda$6584 emission arises in between the
[S~II] zone and the [O~III] zone and can be considered a tracer of
intermediate ionization gas.  The strength of this line is generally
used to estimate the abundance of N$^{+}$ and via the use of an
ionization correction factor, the total nitrogen abundance as well
(e.g.~\cite{mat82}).  Figure \ref{NII} shows the peak [N~II] emissivity
(top panel) and the total [N~II] intensity (bottom panel) for the grid
of models.  As was the case for [S~II], these plots show that the peak
[N~II] is more sensitive indicator of abundance than the total
intensity.  For models using O8, O6 and O4 stellar continua, the [N~II]
intensity decreases with increasing flux.  Also, the [N~II] intensity
decreases for hotter ionizing stars.  The reason is that in both these
cases more nitrogen gets ionized to N$^{++}$ due to the larger number
of energetic photons.  However, the peak emissivity depends on the
N$^{+}$ density in the zone where the emission actually reaches its
maximum.  This ion density is proportional to the total density, which
increases with increasing ionizing flux.  The peak [N~II] emissivity is
therefore higher for greater ionizing flux and for hotter stars.

We have presented a grid of models and discussed the results and
diagnostics obtained from them.  We now consider the specific
case of the blister H~II region M~16 as an example of how these
model results can be applied.

\section{The Blister H~II Region M~16}

HST-WFPC2 images of three elephant trunk structures in M~16 have been
presented and discussed in detail by \cite{hes96}.  Figure
\ref{PCIMAGE} shows the Planetary Camera image of the head of the
second column.  The images have a resolution of 0\farcs046,
corresponding to a linear distance of $1.35\times10^{15}$~cm at an
assumed distance of 2000 pc (e.g. \cite{hil93}).  The H$\alpha$ and
[S~II] images show the ionized interface between the opaque molecular
cloud and the H~II region.  Striations in the H$\alpha$ image are due
to emission from the photoevaporated gas streaming away from the
molecular cloud surface.  The [O~III] emission is much more extended
than the H$\alpha$ and [S~II] emission.  The ratio map of [S~II] to
H$\alpha$ peaks around the edge of the column.  While it may not be
obvious from the image, the ratio along most of the interface is more
or less uniform.

For our analysis and modeling, we will concentrate on the emission
along a spatial cut across the interface at the tip of the column.  The
location is shown in the [S~II] image in Figure \ref{PCIMAGE}.  Line
flux profiles were taken along the cut, and averaged over a width of 5
pixels.  Background intensities were subtracted off, and in addition
the [O~III] profile was smoothed to remove noise.  We note that the
background intensity is mainly due to the back wall of the cavity with
some contribution from the outer parts of the photoevaporative flow.
Here we are concentrating on emission from a narrow region around the
interface itself.  Plots of the H$\alpha$, [S~II] and [O~III] intensity
profiles are shown in Figure \ref{OBSMOD} (top panel).  The intensities
shown in these plots are reddening corrected, taking E$_{B-V}$~=~0.7
(\cite{chi83}) and the ratio of total to selective extinction, R~=~3.1,
the standard value.  For the photoionization model, we assume that the
density profile for the emitting gas is given by equation \ref{EQD} and
from Figure \ref{PCIMAGE}, we find the radius of curvature, $r_{c}$ to
be $2\times10^{17}$~cm.

The peak H$\alpha$ surface brightness is about
$5.5\times10^{-13}$~erg~s$^{-1}$~cm$^{-2}$~arcsec$^{-2}$.  From the
observed geometry (see Figure \ref{PCIMAGE}) we estimate that the path
length through the emitting region lies between about
$6.0\times10^{16}$~cm and $7.5\times10^{16}$~cm.  These values give a
peak H$\alpha$ emissivity lying between about
$4\times10^{-18}$~ergs~s$^{-1}$~cm$^{-3}$ and
$5\times10^{-18}$~ergs~s$^{-1}$~cm$^{-3}$.  The conversion of observed
to intrinsic flux depends on the extinction.  For M~16, \cite{chi83}
have reported that R may be as high as 4.7 (rather than the standard
value of 3.1 which we have so far assumed).  In that case, the
H$\alpha$ emissivity is much higher.  Therefore, from Figure
\ref{HADEN} we estimate that the observed H$\alpha$ emission requires
the density at the interface n$_{\rm H} \gtrsim 4000$ cm$^{-3}$ and
correspondingly an incident ionizing flux $\Phi_{\rm{H}} >
10^{11.0}$~photons~s$^{-1}$~cm$^{-2}$.

The observed [O~III] emission is weak relative to the H$\alpha$
emission (Figure \ref{OBSMOD}, top panel).  By integrating the observed
profiles out to a distance of about $2\times10^{17}$~cm, we estimate
that the total [O~III] intensity is 0.7 times the H$\alpha$ intensity.
The ionizing flux from stars hotter than O6 or cooler than O8 cannot
produce this ratio (Figure \ref{OIII}, bottom panel).  The ratio of the
peak [S~II] emissivity to the peak H$\alpha$ emissivity is between 0.2
and 0.3 (Figure \ref{OBSMOD}, top panel).  From our diagnostic plot
(top panel of Figure \ref{SII}), this implies a sulphur abundance lower
than about 0.4 solar.  Our grid of models uses abundance sets where all
the elements heavier than helium vary in lock-step.  This is a
simplification.  For modeling the interface in M~16 we fixed all
abundances (except for sulphur) based on the values reported by
\cite{haw78}, \cite{sha83} and \cite{den83}.  These values are --
helium solar; nitrogen 0.5 times solar; oxygen and the other metals 0.8
times solar.  The sulphur abundance was allowed to vary.

The H$\alpha$, [S~II] and [O~III] profiles, along with the density
profile for our final M~16 model are presented in the bottom panel
of Figure \ref{OBSMOD}.  The ionizing continuum used in the model
is from an O7Ia supergiant, with T$_{eff} = 38720$~K and the ionizing
flux at the interface is $3\times10^{11}$~photons~s$^{-1}$~cm$^{-2}$.
The sulphur abundance is 0.4 times solar.  The photoionization 
model reproduces the observed emission from the interface, including
the stratified ionization structure of the flow.  In matching the
model to the observation, we have assumed a plane-parallel geometry --
clearly reality is more complex (indeed we have assumed a scale
length for the flow based on the radius of curvature of the column).
However since we are concentrating on the emission from close to
the interface, for our purposes the plane parallel approximation is
sufficient.

In Table \ref{TBL_SPEC} we present the observed (reddening-corrected)
spectrum of M~16 measured by \cite{haw78} along with the spectrum
predicted by our final model.  All intensities are given relative to
H$\alpha$~=~100.  The two observed spectra were taken along slits
2\farcs4 by 4\farcs0 separated by 35\arcsec\ along an east-west line in
a region at the top of column 1 of the HST image presented by
\cite{hes96}.  The spectra at the observed positions vary
significantly.  The [S~II]~$\lambda\lambda$ 6716,6731 for example is
higher by a factor of 4 at position 2.  The [O~III]~$\lambda$5007 on
the contrary is lower by a factor of 2/3.  Our model spectrum matches
the position 1 spectrum rather well.  The major exception is the
[O~II]~$\lambda$3727 line, where the mismatch may be due to a
repositioning error since the [O~II] line was taken through a separate
grating (see \cite{haw78}).  Another difference is in the ratio between
the two [S~II] lines.  Since the exact location of the observed
spectrum is not known and since the observation and model pertain to two
different regions (columns 1 and 2 respectively),  the reason for the
difference is not clear.  The total strength of the two lines (relative
to H$\alpha$) however is reproduced by the model.  It is well known
that spectra vary within an H~II region.  The significance here is that
our model matches one of the spectra so well.  This allows us to infer
that the major reason for the differences in the two observed spectra
are due to different orientations of the slit with respect to the
highly stratified emitting interface.

\section{Concluding Remarks}

In this work, we have presented photoionization models for the emission
from the photoevaporative interface between the ionized gas and the
molecular cloud in blister H~II regions.  The density profile of the
interface is a power-law which can be parameterized by the distance
from the ionizing source to the ionization front and the density at the
ionization front (\cite{ber96}).  The values of these parameters depend
upon the shape and strength of the ionizing continuum.  

From the grid of models, which have systematically varying input
parameters, we find that the H$\alpha$ peak emissivity is strongly
dependent on the incident ionizing flux, increasing for higher values
of flux.  We also find that the [O~III] emission is most sensitive to
the type of star responsible for ionizing the interface.  Stars cooler
than O8 produce photoionized interfaces with virtually no [O~III]
emission.  The low ionization [S~II] emission is confined to a very
sharp zone.  We find that the peak [S~II] emissivity is more sensitive
than the total [S~II] intensity to the sulphur abundance.  The same
holds true for the intermediate ionization [N~II] line.  It is worth
noting that in the [S~II] zone, the [S~II] $\lambda\lambda$6716,6731
to H$\alpha$ ratio can get relatively high (e.g.~it is $> 0.5$ for the
model shown in the bottom panel of Figure \ref{CSTAR}).  Such high
ratios are often considered to be a signature of shock-excited gas
but our models show that they can also occur in photoionized interfaces.

We then considered narrow band HST data of the blister H~II region
M~16.  The emission and ionization structure seen at high resolution
was used to constrain the properties of the ionizing continuum and the
sulphur abundance.  We presented a specific model that reproduced the
observations in detail.  We found that the integrated spectrum
predicted by the model matched one of two ground based spectra taken by
\cite{haw78} at a nearby location.  These data also validate our
photoionization models.  Our being able to match the observed emission
and ionization structure with these models is strong evidence that
blister H~II regions can be described by a photoionized,
photoevaporative flow.

The method we have presented here is important for the study of H~II
regions, both in our Galaxy and in other galaxies.  In the case of
sufficiently nearby objects (such as M~16) we can obtain spatially
resolved spectra.  Then, our current study indicates that we could use
photoionization modeling to obtain elemental abundances without having
to make assumptions about the structure of the emitting region.
Carrying out such an exercise for a large sample of H~II regions will
be a crucial test for the applicability of the model for a range of
conditions.  We can apply our understanding of the emission from nearby
H~II regions to more distant H~II regions (governed by the same
physical mechanisms).  Specifically, we can calculate photoionization
models of photoevaporative flows in order to interpret spectra of these
objects.  One important class of distant emission nebulae which we may
be able to study within this framework are the giant extra-galactic
H~II regions (GEHRs).  It is promising that narrow band HST images of
30 Doradus, the nearest GEHR, show the optical emission
concentrated in sharp photoevaporative interfaces, much like in M~16
(\cite{sco98}).  A direct application of our method to study the
conditions in 30 Doradus would be an important step towards extrapolating
to GEHRs in other galaxies.

\acknowledgements

RS thanks Jason Aufdenberg for discussions about stellar atmospheres
and for providing the PHOENIX models used in our calculations.  We
thank the anonymous referee for several useful comments.  Part of this
work was completed while RS was a graduate student at Arizona State
University.  This work was partly supported by NASA/JPL contracts
959289 and 959329.


\clearpage

\figcaption{The plot shows the PHOENIX model spectra of B0, O8, O6 and
O4 stars.  These have been used as incident continua in the grid of
models presented here.  The ionization potentials for S$^{0}$ and
O$^{+}$ are shown.  The edge in each of the spectra is at 13.6 eV,
corresponding to the ionization potential for H$^{0}$.  Note that the
fraction of photons with energies greater than 35.12 eV decreases with
increasing stellar temperature, while the fraction of photons with
energies between 10.4 eV and 13.6 eV increases.
              \label{CONTINUA}}

\figcaption{Each plot shows model predictions of spatial profiles of
the density, and the H$\alpha$, [S~II] and [O~III] emission.  The input
parameters for the two models shown differ only in the incident
ionizing flux, $\Phi_{\rm{H}}$, values of which are shown on the
plots.  The model with the higher incident flux (bottom panel) leads to
a higher density at the interface, correspondingly stronger line fluxes
and a narrower [S~II] emission zone.
              \label{CPHI}}

\figcaption{Plots are shown as in Figure \protect\ref{CPHI} for models
differing in the ionizing stellar atmosphere used.  The hotter star (bottom
panel) leads to significantly higher [O~III] emission and also a narrower
[S~II] emitting zone.
              \label{CSTAR}} 

\figcaption{Plots are shown as in Figure \protect\ref{CPHI} for models with
different elemental abundances.  The increased sulphur abundance leads not
only to a higher [S~II] line intensity, but also a greater contrast in the
ratio of peak [S~II] emissivity to the peak H$\alpha$ emissivity.
              \label{CABUND}} 

\figcaption{The plot compares the PHOENIX O4V (T$_{eff}$ = 48,670~K)
spectrum with the CoStar E2 (T$_{eff}$ = 48,500~K) spectrum
spectrum.  The ionization potentials for S$^{0}$ and O$^{+}$ are shown.
The CoStar model includes the effects of a wind which results in a
flatter spectrum and consequently a higher fraction of photons able to
produce O$^{++}$.
            \label{COS_PHX_CONT}}

\figcaption{Plots are shown as in Figure \protect\ref{CPHI} for models
using PHOENIX and CoStar incident spectra.  The incident ionizing flux
and abundances are the same for both models.  The main difference is that
the CoStar model predicts twice as much [O~III] as the PHOENIX model.
            \label{COS_PHX_SPEC}}

\figcaption{Top Panel: The plot shows the densities at the ionization
edge for the grid of models.  Different symbols stand for models with
different ionizing flux.  The plot is divided into four broad regions
along the x-axis, each with models using a particular stellar ionizing
continuum.  Within each group of five consecutive models, the abundance
of metals heavier than helium varies from 0.2 to 1.0 in steps of 0.2
going from left to right.  The strong dependence of maximum density
reached on incident ionizing flux is clearly seen.  (The organization
of this plot is used for the plots shown in Figures \protect\ref{OIII},
\protect\ref{SII} and \protect\ref{NII}).  Bottom Panel: The scatter
plot of peak H$\alpha$ emissivity vs. maximum density reached shows the
strong correlation between these two quantities.  The maximum density
in turn is most sensitive to the ionizing flux.
                \label{HADEN}}

\figcaption{The peak [O~III] emissivity and the total [O~III]
intensities for the grid of models are shown in the top and bottom
panels respectively.  
                \label{OIII}}

\figcaption{The peak [S~II] emissivity and total [S~II] intensities for
the grid of models are shown in the top and bottom panels.  For a given
ionizing continuum, the peak emissivity is a much more sensitive
indicator of abundance than the total intensity.
                \label{SII}}

\figcaption{The peak [N~II] emissivity and total [N~II] intensities for
the grid of models are shown in the top and bottom panels.  As in the
case of [S~II], the peak emissivities are more sensitive indicators of
abundance than total intensities.
                \label{NII}}

\figcaption{HST images of the top of a ``pillar'' in M~16 showing the
field of view of the PC.  The brightened rim around the column is seen
clearly in the H$\alpha$ and [S~II] images, and also stands out in the
[S~II] to H$\alpha$ ratio map.  The ratio is more or less uniform all
around the edge.  The ionizing stars lie directly above the tip of the
column.  The location of the cut across the interface for which
emission profiles have been analyzed is shown in the [S~II] image.
             \label{PCIMAGE}}

\figcaption{Top Panel:  The observed profile across the photoionized
interface.  The [O~III] profile has been smoothed for clarity.  The
distance scale assumes a distance of 2000 pc to the nebula, and the
peak of the [S~II] emission lies at 0.  Bottom Panel: The emission
profiles predicted by the photoionization model.  The observed
intensity is directly proportional to the model emissivity so the two
plots can be directly compared.  The density profile used in the
model is also shown.
             \label{OBSMOD}}

\clearpage

\begin{deluxetable}{ccccc}
\tablecaption{Model line strengths compared with observations. 
                                   \label{TBL_SPEC}}
\tablewidth{0pt}
\tablehead{
  \colhead{Line} & \colhead{$\lambda$} & \colhead{Pos. 1}  
                 & \colhead{Pos. 2}    & \colhead{Model}    
}
\startdata
 H$\alpha$   & 6563 & 100.0   &  100.0   &  100.0     \nl
     He~I    & 5876 &   4.1   &    6.3   &    4.7     \nl
   {[N~II]}  & 5755 &   0.1   &    0.8   &    0.2     \nl
   {[N~II]}  & 6548 &   5.7   &   17.5   &    7.6     \nl
   {[N~II]}  & 6584 &  18.2   &   55.4   &   22.5     \nl
   {[O~II]}  & 3727 &  56.1   &  114.6   &  109.3     \nl
  {[O~III]}  & 4363 & $<$1.4  & $<$1.5   &    0.2     \nl
  {[O~III]}  & 4959 &  16.4   &    9.8   &   22.8     \nl
  {[O~III]}  & 5007 &  50.5   &   32.3   &   66.0     \nl
   {[S~II]}  & 6716 &   2.8   &    8.0   &    2.1     \nl
   {[S~II]}  & 6731 &   2.6   &   10.5   &    3.4     \nl
  {[S~III]}  & 6312 & $<$0.1  &    0.8   &    0.3     \nl
\enddata
\tablecomments{The observed values are taken from \protect\cite{haw78}}
\end{deluxetable}

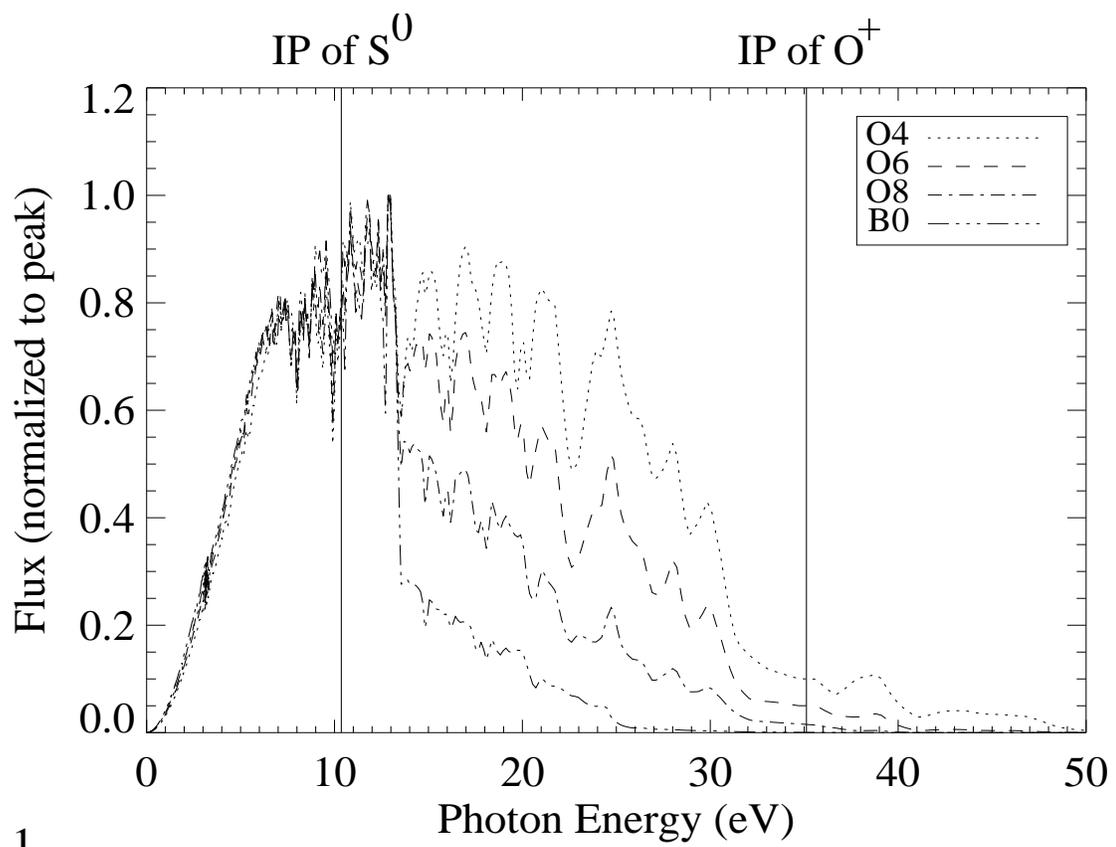
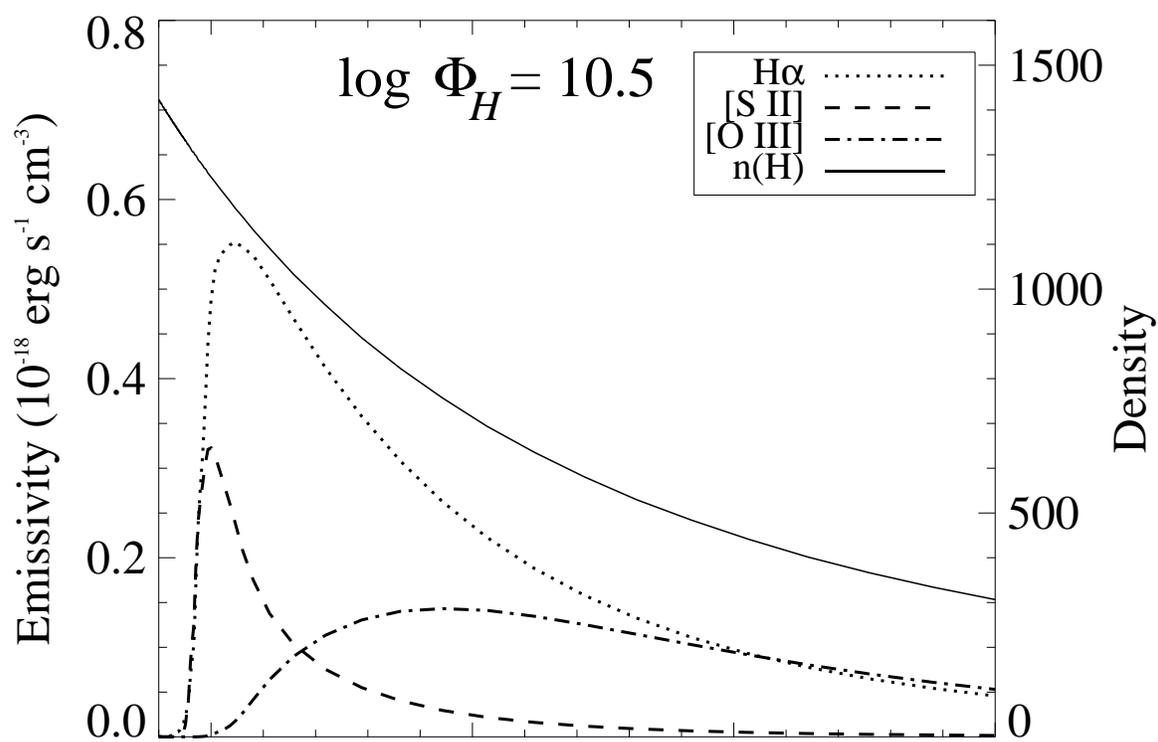
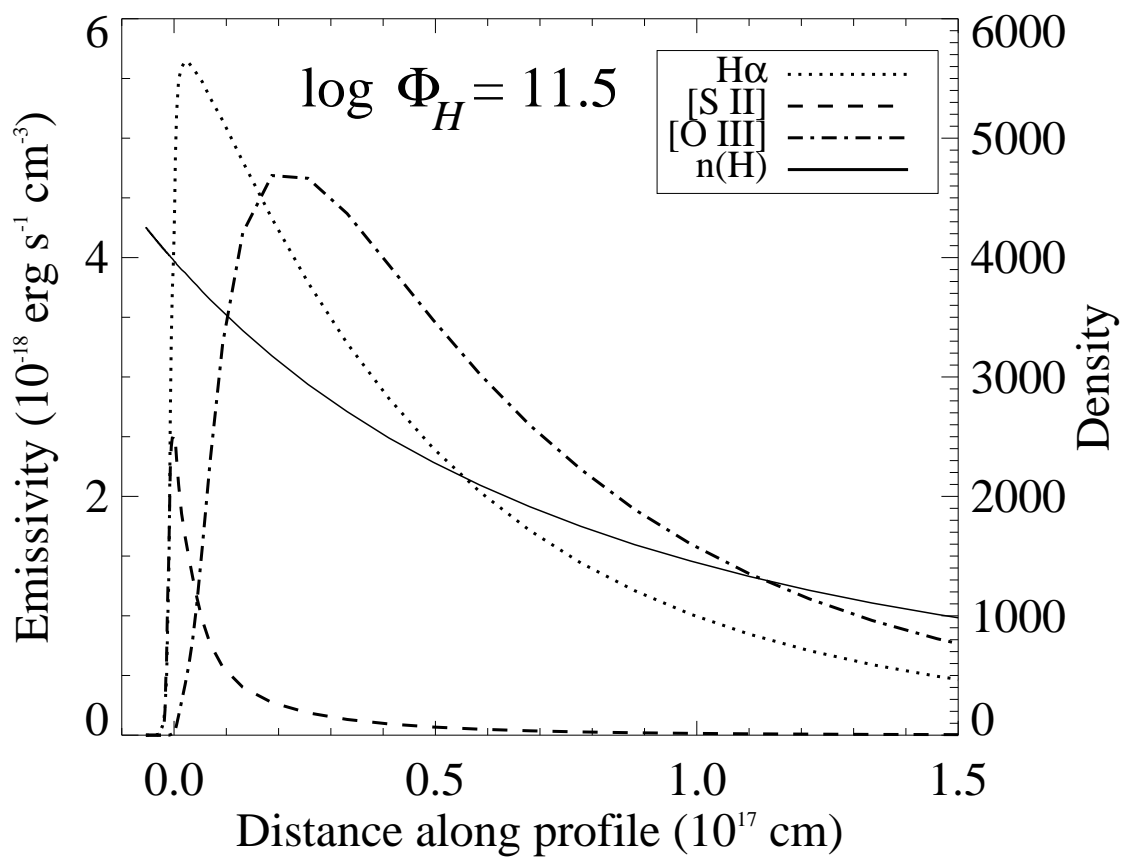
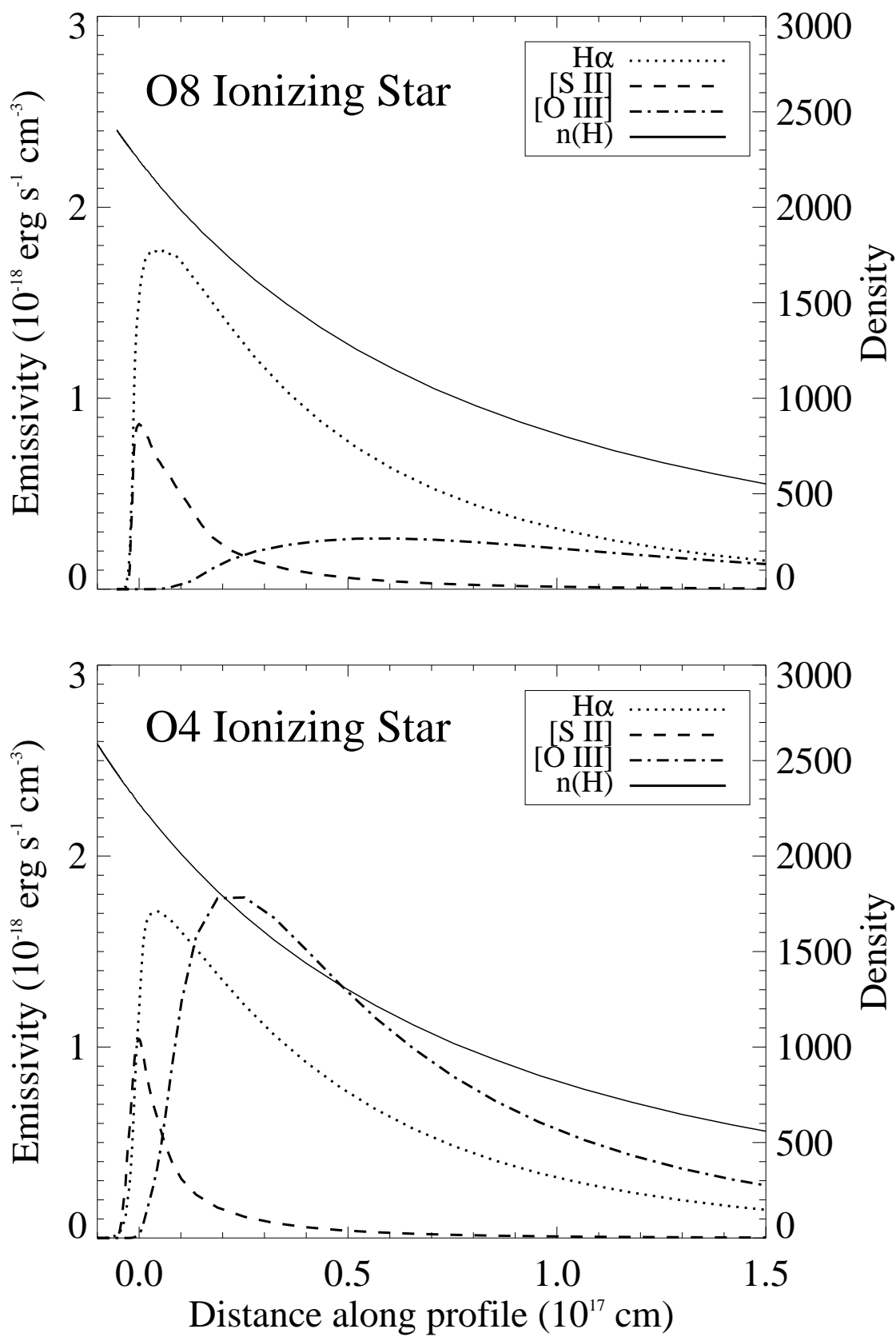
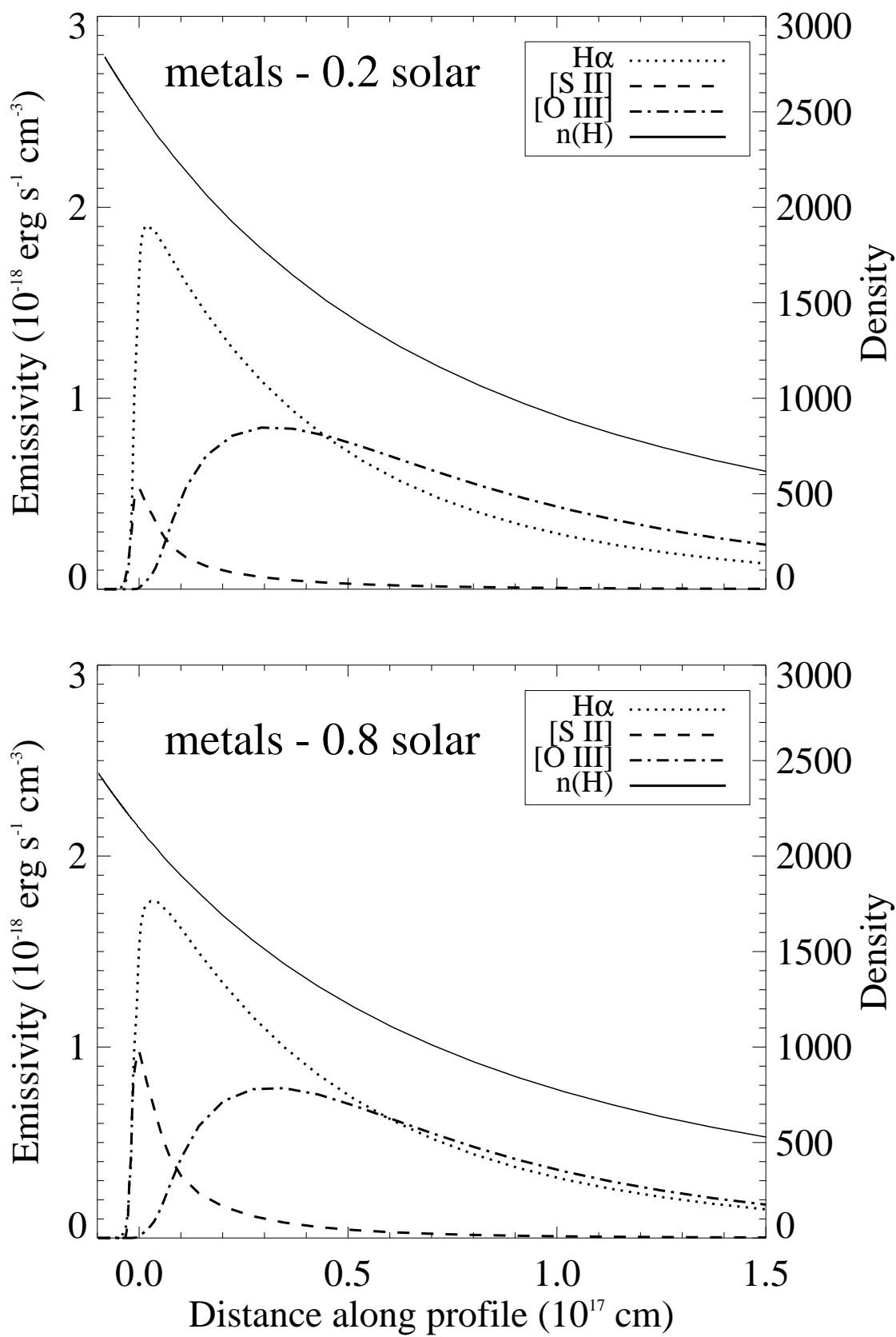
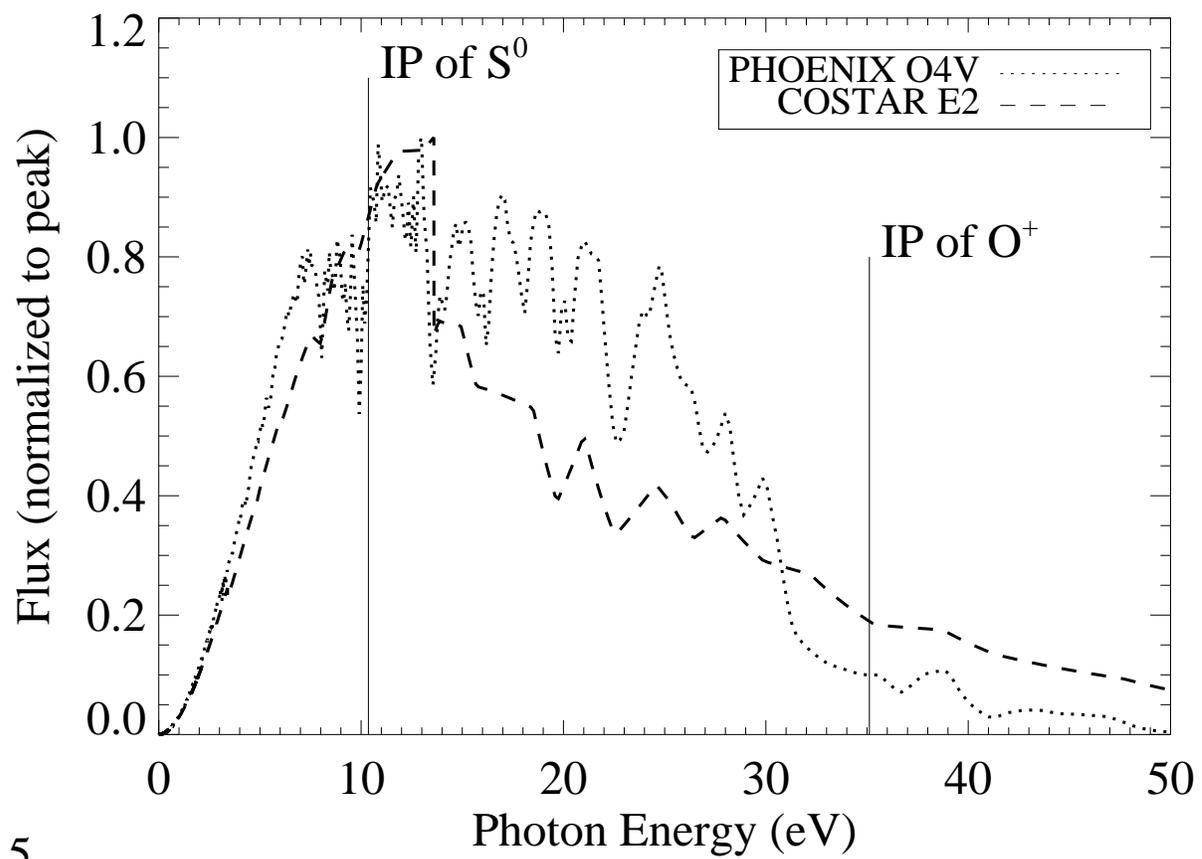
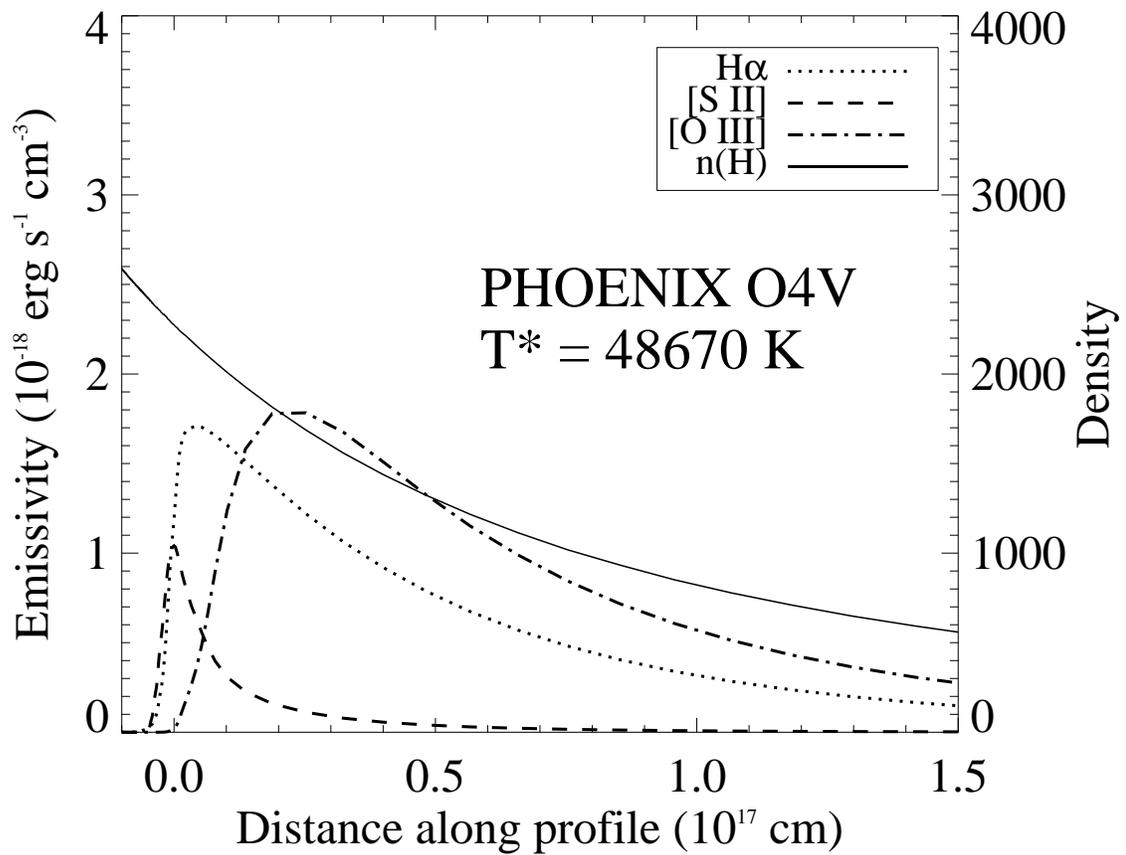
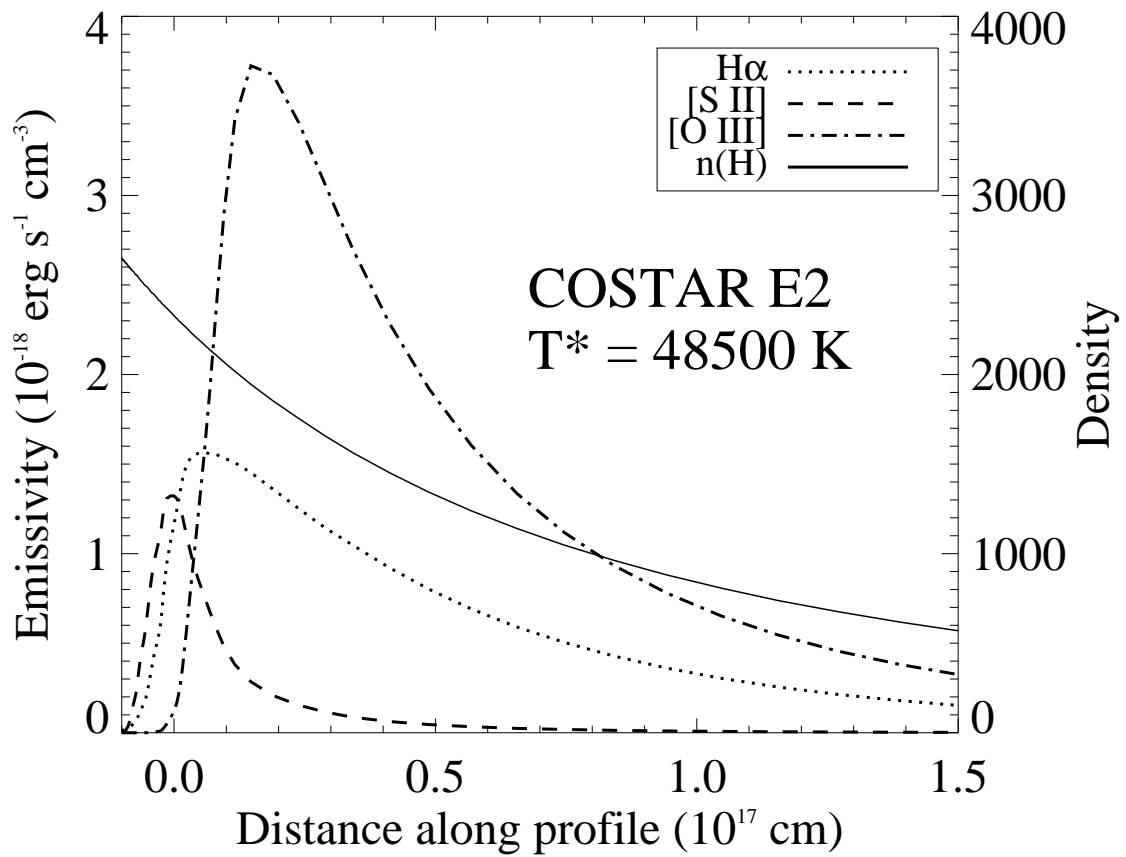
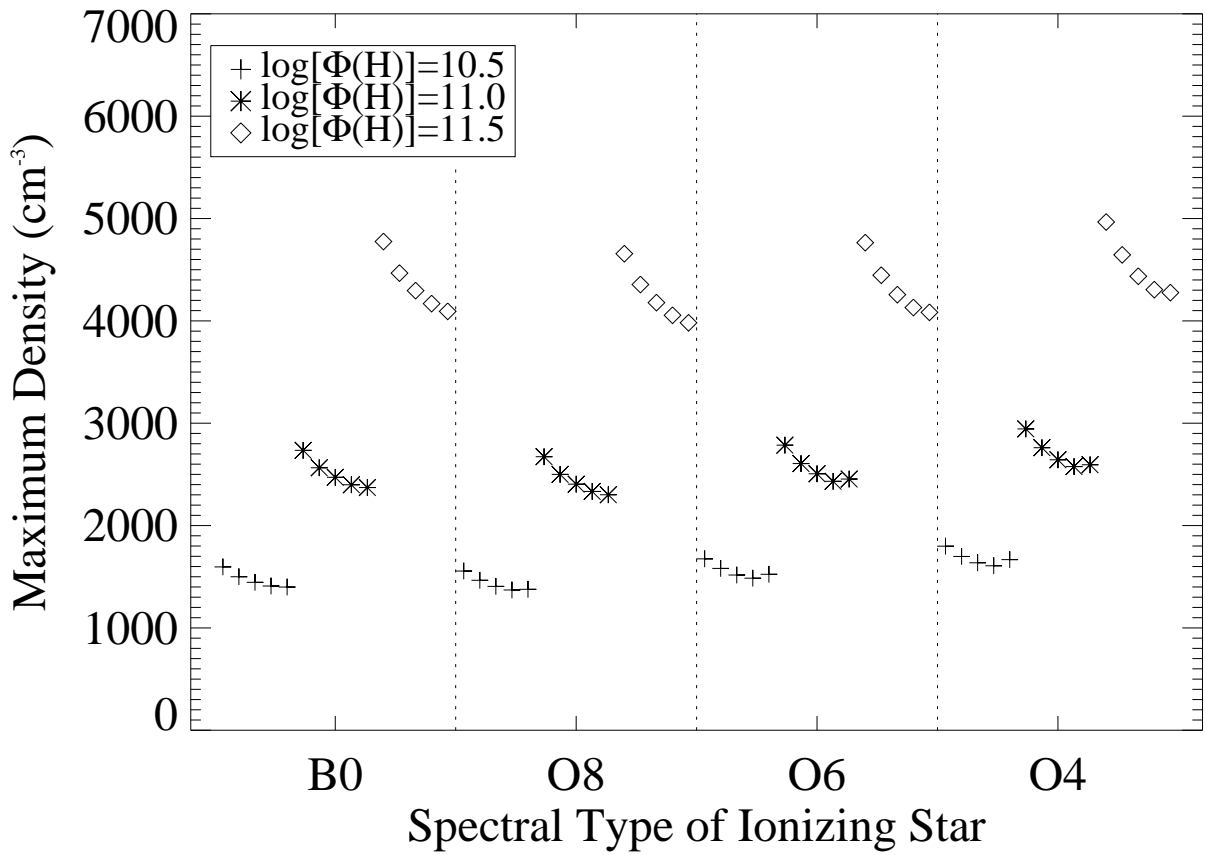
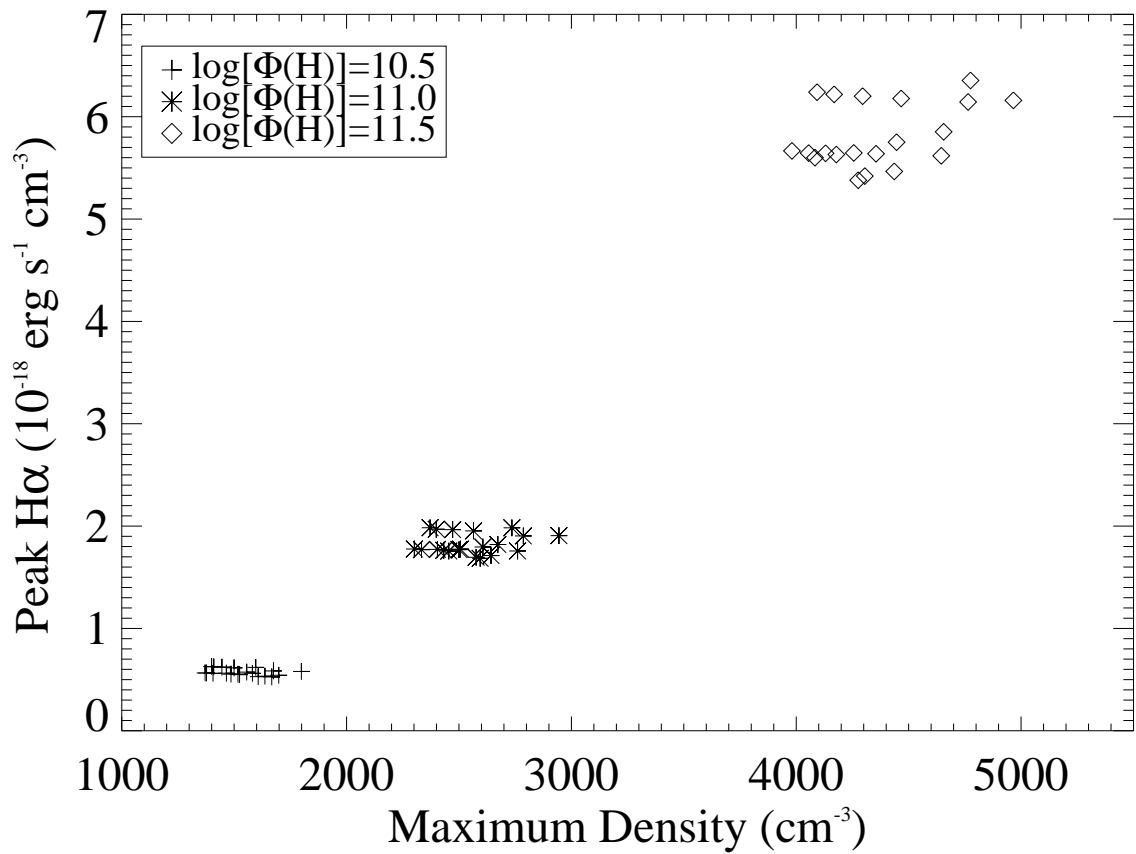
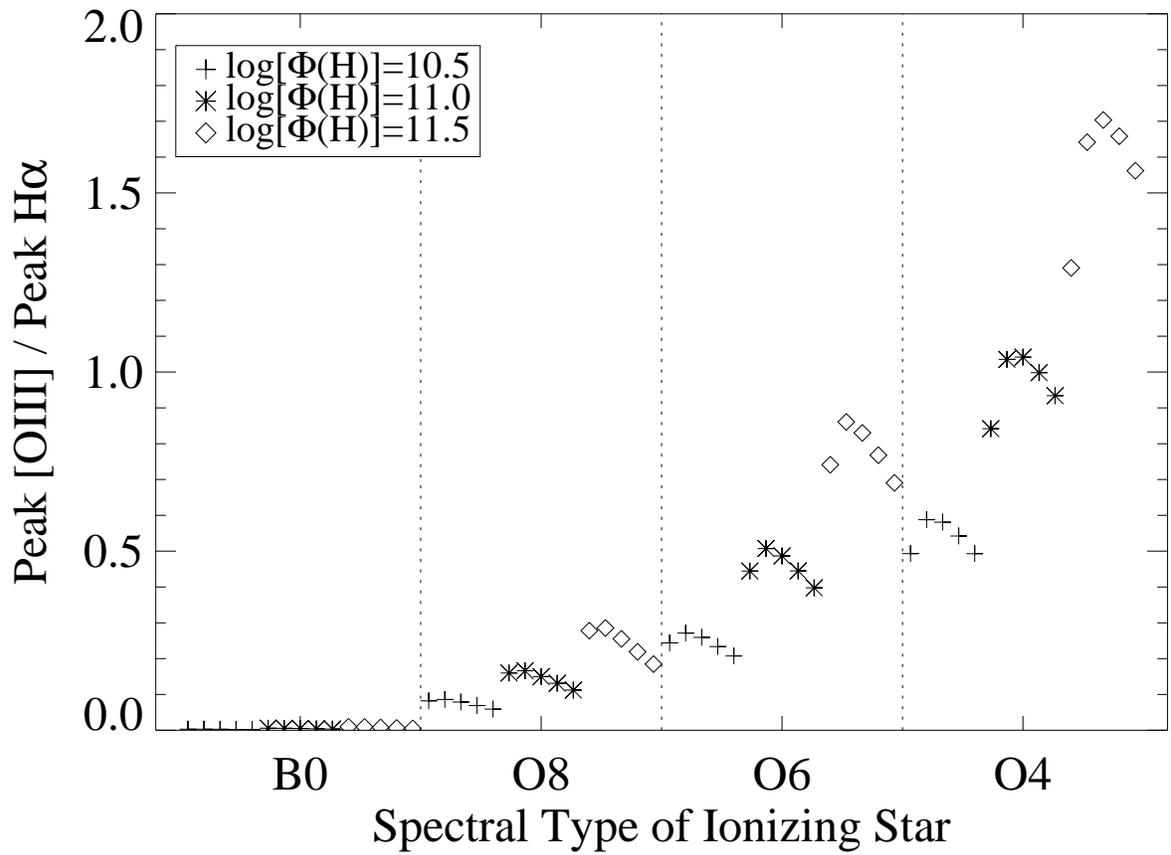
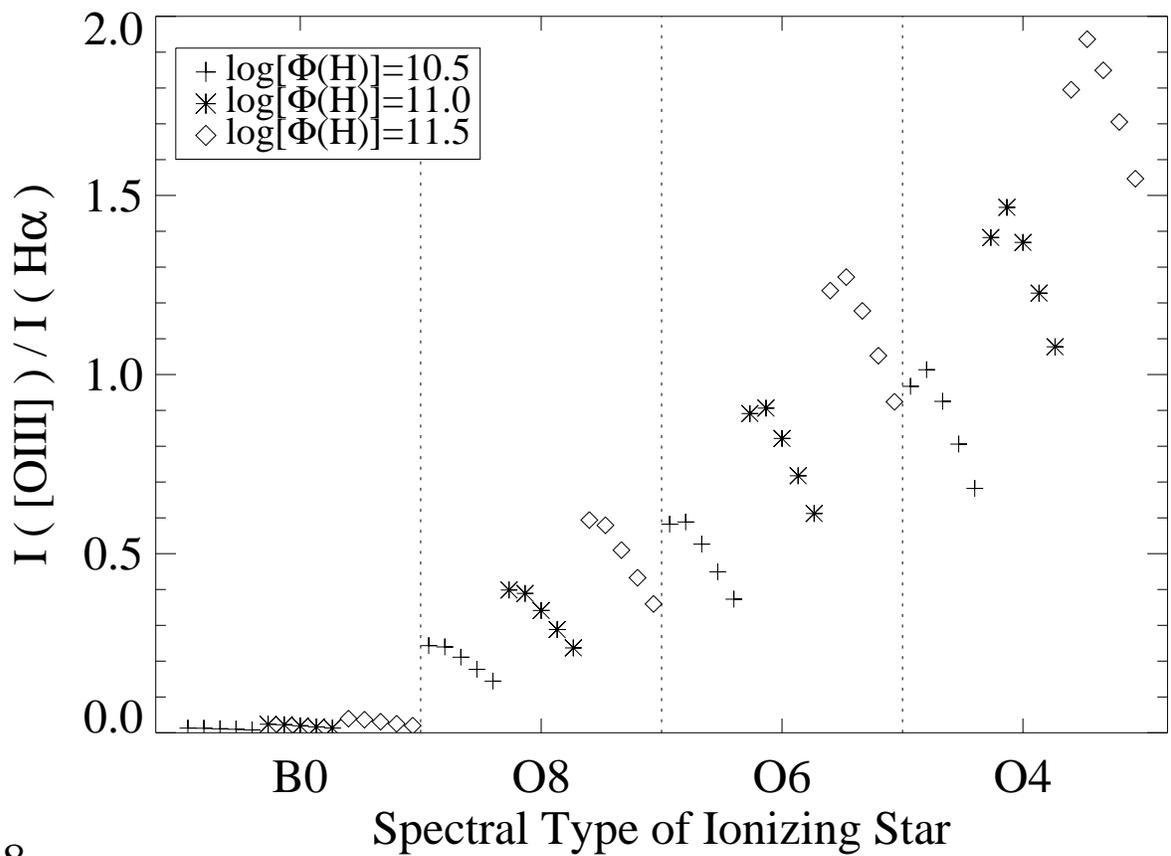
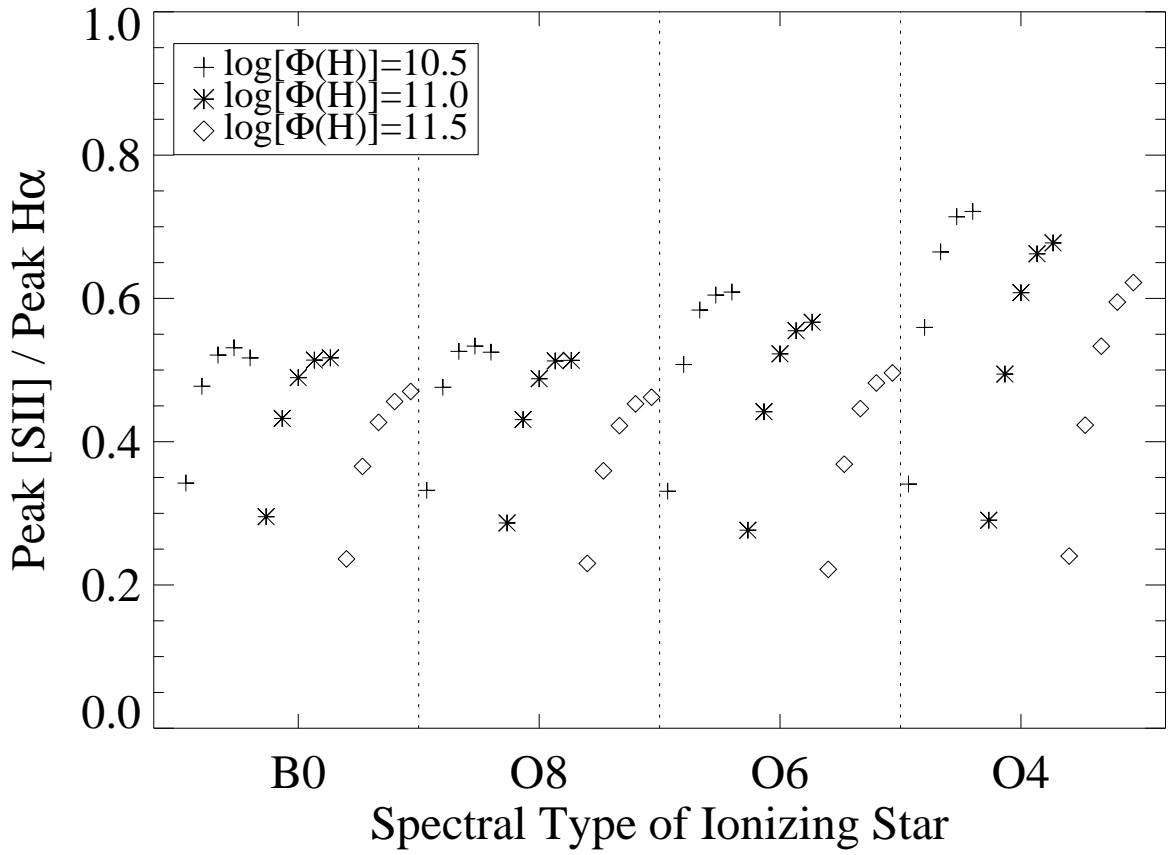
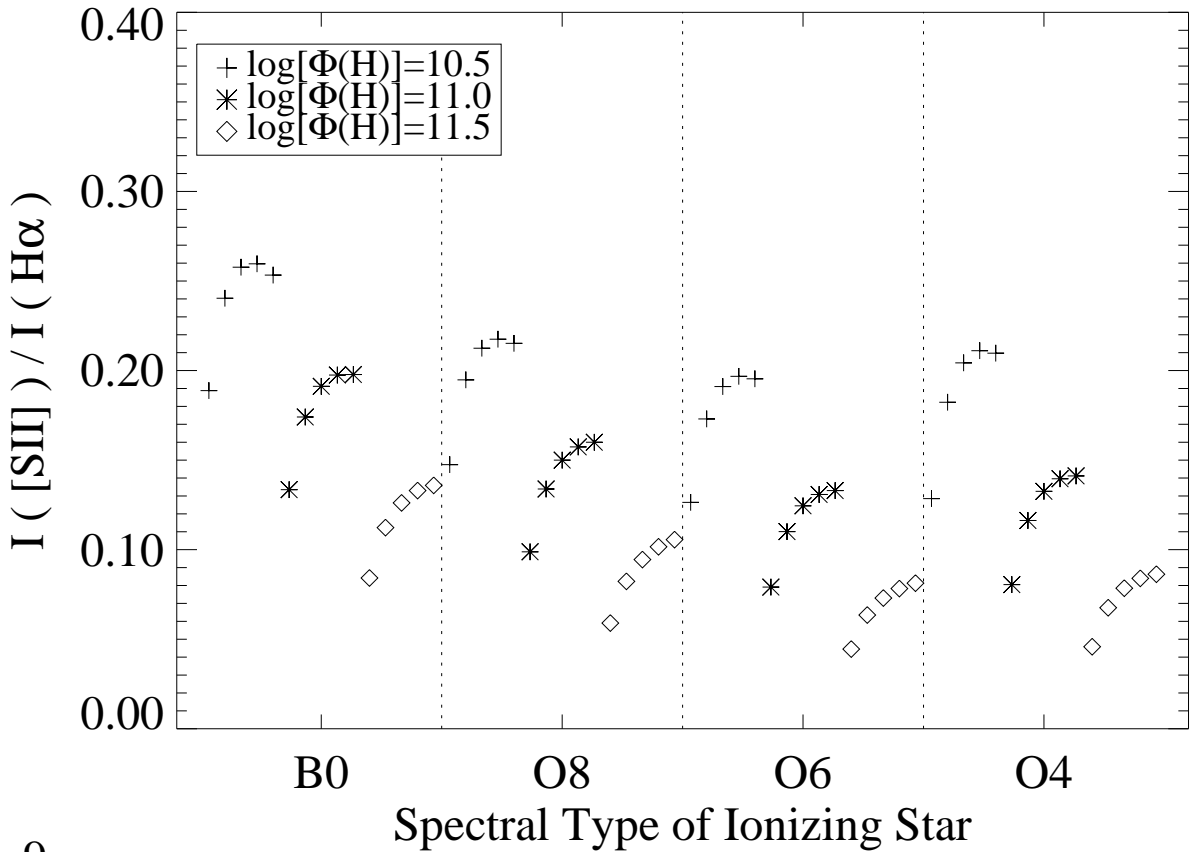
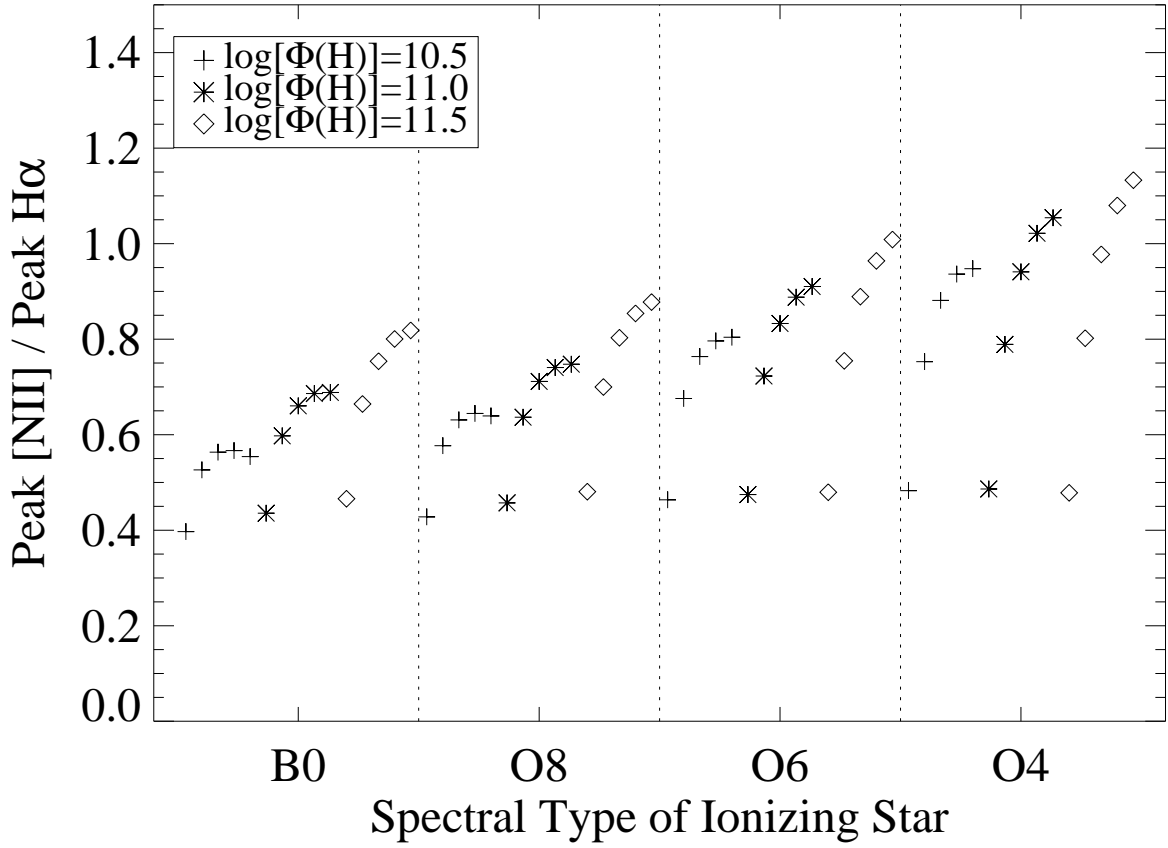
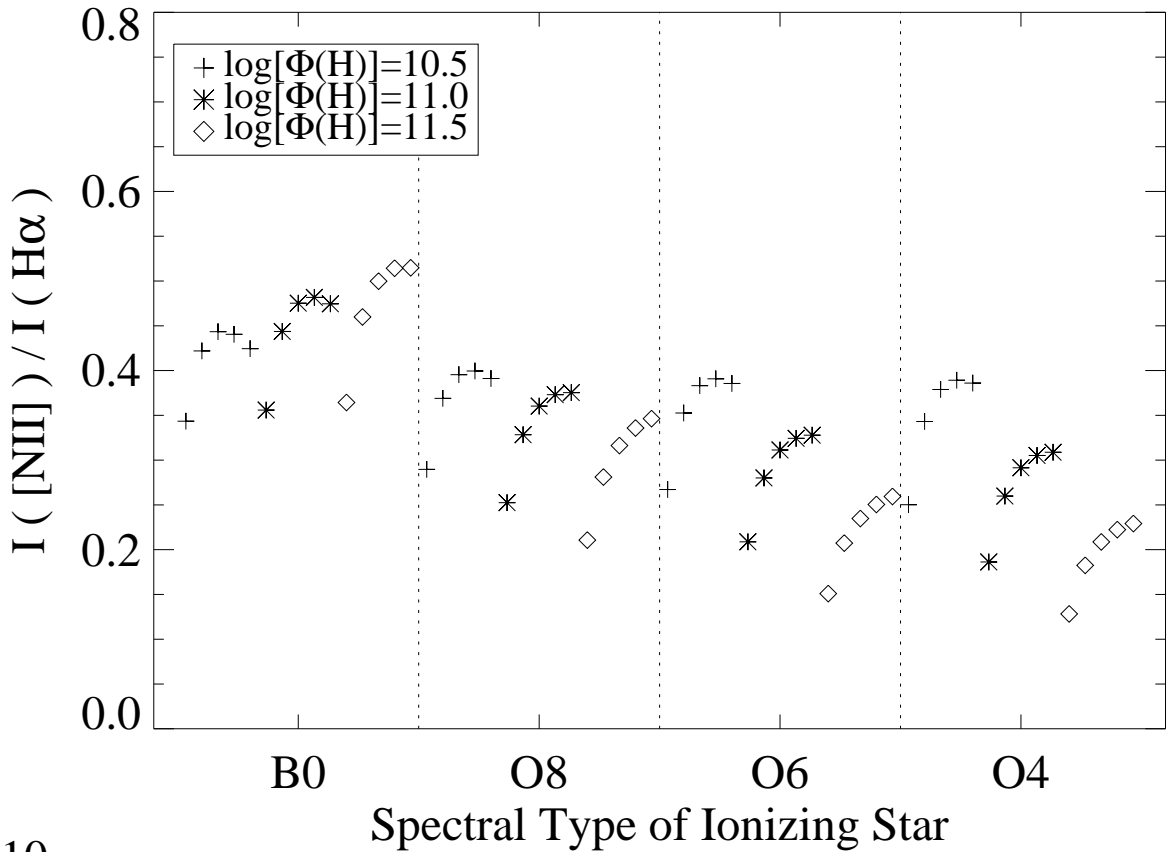
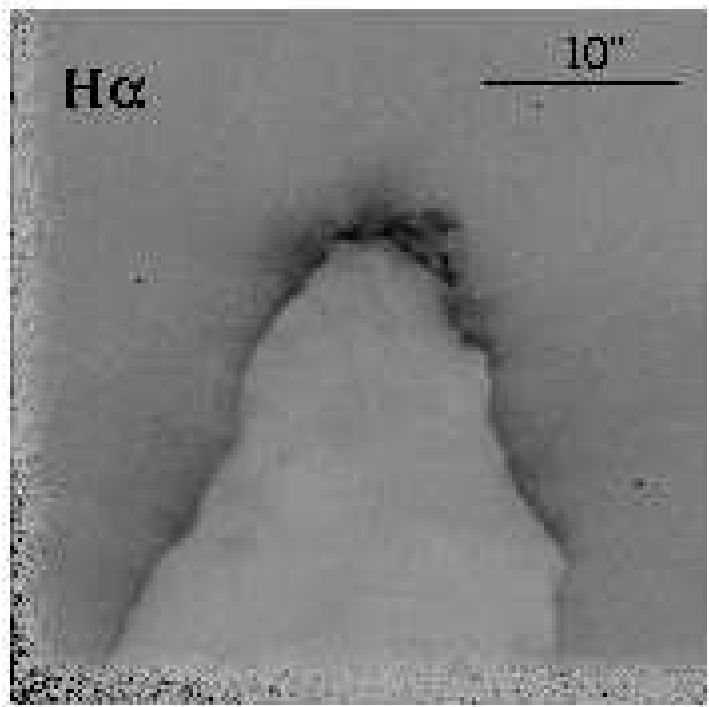
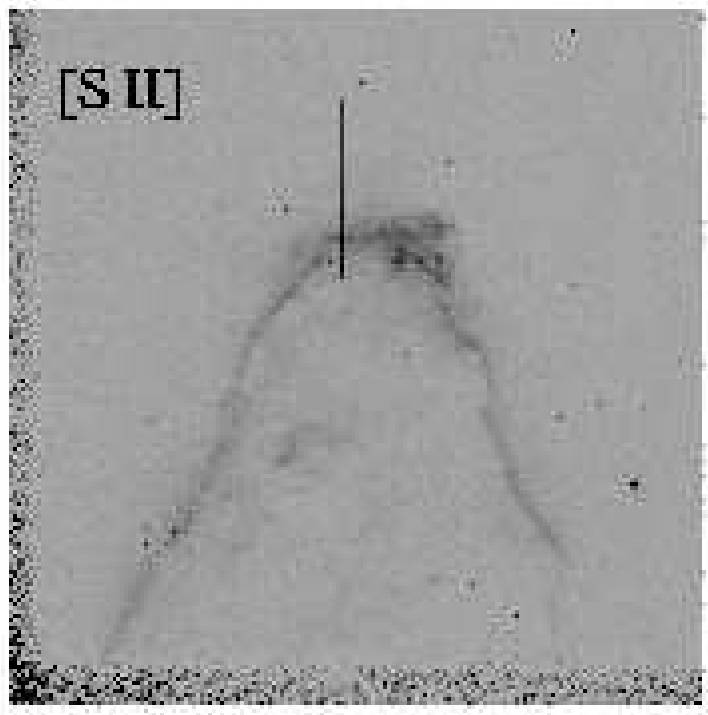
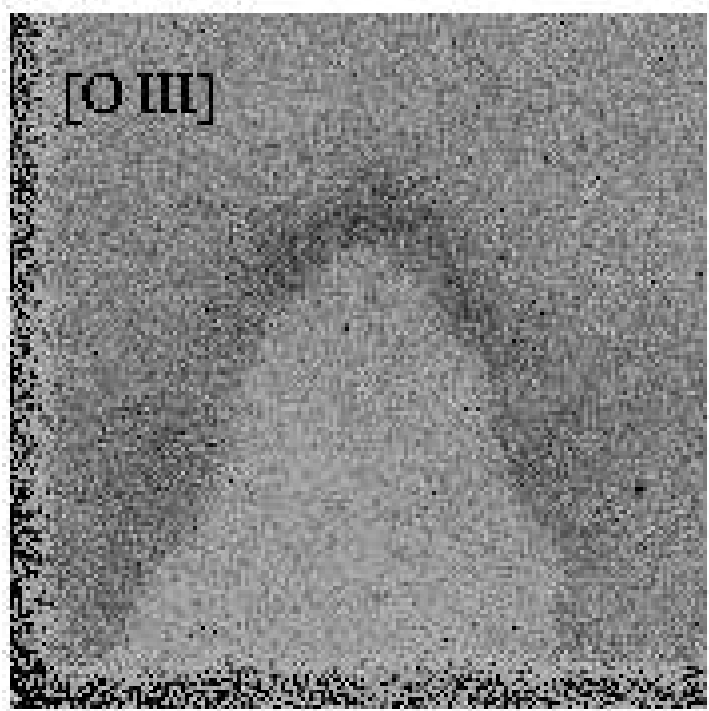
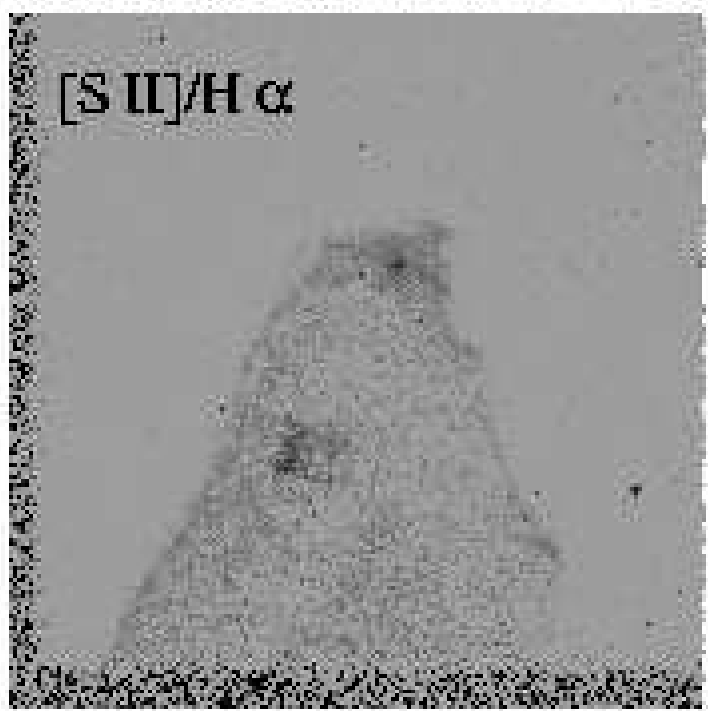
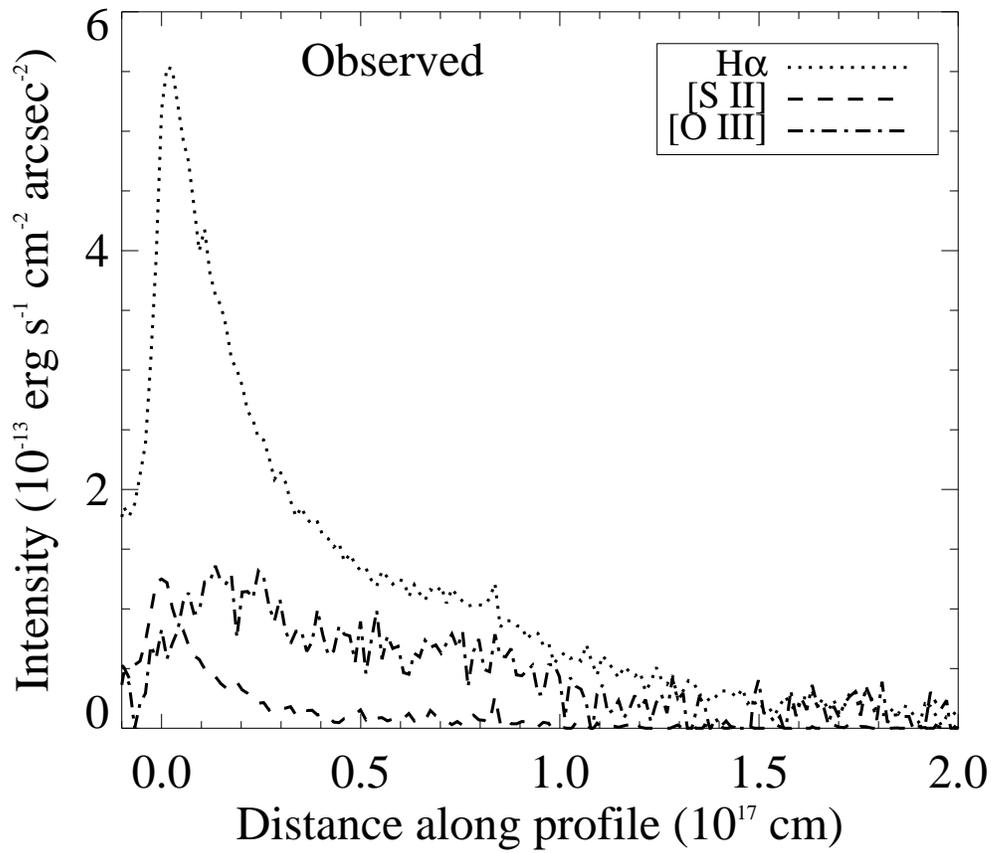
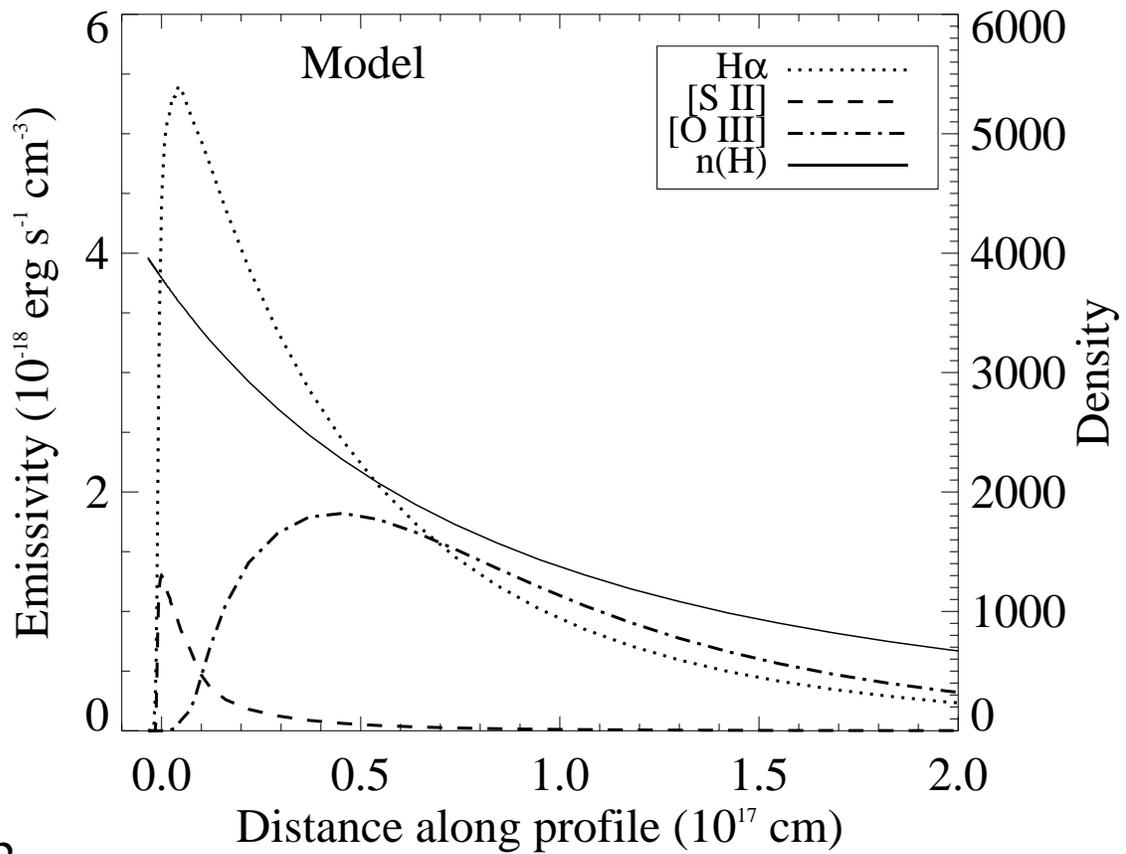
\end{document}